# Tunable superconducting resonators via on-chip control of local magnetic field

Chen-Guang Wang(王晨光)[1,2], Wen-Cheng Yue(岳文诚)[1], Xuecou Tu(涂学凑)[1], Tianyuan Chi(迟天圆)[1], Tingting Guo(郭婷婷)[1], Yang-Yang Lyu(吕阳阳)[1], Sining Dong(董思宁)[1], Chunhai Cao(曹春海)[1], Labao Zhang(张蜡宝)[1,3], Xiaoqing Jia(贾小氢)[1,3], Guozhu Sun(孙国柱)[1,3], Lin Kang(康琳)[1,3], Jian Chen(陈健)[1,2], Yong-Lei Wang(王永磊)[1,2,4,†], Huabing Wang(王华兵)[1,2,†], and Peiheng Wu(吴培亨)[1,2,3]

[1] *Research Institute of Superconductor Electronics, School of Electronic Science and Engineering, Nanjing University, Nanjing 210023, China*
[2] *Purple Mountain Laboratories 211111, Nanjing, China*
[3] *Hefei National Laboratory, Hefei 230094, China*
[4] *State Key Laboratory of Spintronics Devices and Technologies, Nanjing University, Nanjing 210023, China*

Superconducting microwave resonators play a pivotal role in superconducting quantum circuits. The ability to fine-tune their resonant frequencies provides enhanced control and flexibility. Here, we introduce a frequency-tunable superconducting coplanar waveguide resonator. By applying electrical currents through specifically designed ground wires, we achieve the generation and control of a localized magnetic field on the central line of the resonator, enabling continuous tuning of its resonant frequency. We demonstrate a frequency tuning range of 54.85 MHz in a 6.21 GHz resonator. This integrated and tunable resonator holds great potential as a dynamically tunable filter and as a key component of communication buses and memory elements in superconducting quantum computing.

## 1. Introduction

Superconducting microwave resonators are key elements in applications such as photon detection [1] and quantum computing [2], owing to their simplicity in fabrication, ease of integration, and low energy loss. The demand for superconducting resonators with in-situ tunable resonance frequencies has grown across diverse applications, including parametric amplifiers [3-6], spectrometers [7], magnetometers [8] and microwave storage devices [9]. The frequency tunability can be achieved by tuning the nonlinear inductance of a superconducting quantum interference device (SQUID) [3, 10-12]. While these SQUID-based resonators usually exhibit substantial frequency tuning range due to the considerable nonlinear inductance of Josephson junctions, the relatively low critical currents of Josephson junctions prevent their applications with high operating power [6] or in strong magnetic fields [18]. In contrast, the resonant frequency of superconducting resonators can be tuned by altering the kinetic inductance of superconductors [13-19]. This can be achieved by injecting a DC current into the resonator from an integrated DC bias circuit [13-15]. However, this unavoidably introduces additional losses to resonators. This limitation can be overcome by tuning the kinetic inductance of superconductors using magnetic fields, leading to remote tuning of superconducting resonators without the need for adding additional circuits into the resonator line [16-19]. However, in all previously demonstrated magnetic-field-tunable superconducting resonators, the magnetic field is applied externally to the entire chip. This not only hiders applications in magnetic field sensitive systems [20-24] but also limits their potential for large-scale integration. In this work, we address these limitations with a unique approach, on-chip generating a local magnetic field to control the resonance frequency of a superconducting coplanar waveguide (CPW) resonator. This is achieved by applying anti-parallel supercurrents through two meticulously designed ground lines, which effectively confines the magnetic field to the resonator's central line, thus presenting notable advantage for system integration.

† Corresponding author. E-mail: yongleiwang@nju.edu.cn; hbwang@nju.edu.cn

## 2. Device fabrication and parameters

The kinetic inductance of a superconductor is proportional to the superconducting penetration depth [25]. Niobium nitride (NbN) is known for its long penetration depth, ranging from 300 nm to 400 nm [26]. Thus, in this study, we utilize NbN films to fabricate the superconducting resonator. As illustrated in Fig. 1, our device consists of a half-wavelength CPW resonator capacitively coupled to a feedline. First, a 22 nm thick NbN film is sputtered onto a C-plane sapphire substrate, and then the device structure is formed using standard photolithography followed by $CF_4$ dry etching. Two current ground lines are separated from the ground plane, each with one side directly connected to the ground plane and the other side linked to DC sources. To transmit higher supercurrent through the ground wires for generating higher magnetic field, we increase the thickness of the two ground wires with 270 nm thick NbN film, achieved through the standard photolithography and lift-off techniques. Because magnetic field produces superconducting vortices, the quantized magnetic fluxes, in the center conductor, to avoid the vortices' motion induced energy losses, we introduce holes in the center conductor to pin the vortices [27-30]. This also generates circulating supercurrents around the holes when magnetic field is applied, thus altering the kinetic inductance of the center line [16].

The geometric inductance $L_g$ and the capacitor $C$ of the resonator can be calculated using conformal mapping [31]. They are given by

$$L_g = l \cdot \frac{\mu_0}{4} \frac{K(k_0')}{K(k_0)}, \qquad (1)$$

$$C = l \cdot 4\epsilon_0 \epsilon_{eff} \frac{K(k_0')}{K(k_0)}, \qquad (2)$$

where $l$ ($\approx 4.3$ mm) is the length of the CPW resonator, and $\epsilon_{eff}$ ($\approx 6.3$) is the effective permittivity of the CPW line. Here, $K$ denotes the complete elliptic integral of the fist kind with the arguments

$$k_0 = \frac{w}{w + 2s}, \quad (3)$$

$$k_0' = \sqrt{1 - k_0^2}, \quad (4)$$

where $w$ is the width of the center conductor (10 μm), $s$ is the gap between the center conductor and the ground wires (3 μm). Hence, the geometric inductance $L_g$ is 1.509 nH, and the capacitor $C$ is 0.854 pF. The fundamental mode $f_0$ of a half-wavelength resonator is given by $f_0 = 1/2\sqrt{(L_g + L_{k0})C}$, where $L_{k0}$ denotes kinetic inductance [31]. The experimentally determined $f_0$ is around 6.21 GHz. Thus, we can determine the kinetic inductance $L_{k0}$=6.079 nH.

Our device exhibits a high kinetic inductance ratio $\alpha = L_{k0}/(L_g + L_{k0}) \approx 0.801$, which is advantageous for tuning the resonant frequency by changing the kinetic inductance. The kinetic inductance can be tuned by magnetic field. The relation between the kinetic inductance of a superconductor and the magnetic field $H$ is given by [2, 19]

$$L_k(H) \approx L_{k0}\left(1 + \frac{H^2}{H_*^2}\right), \quad (5)$$

where $L_{k0}$ is the kinetic inductance in zero magnetic field, and $H_*$ is characteristic magnetic field related to critical magnetic field. $L_k$ typically exhibits limited tunability due to $H \ll H_*$ [2]. The resonance frequency shift ratio is given by

$$\frac{\Delta f}{f_0} = -\beta H^2, \quad (6)$$

where $\beta = \alpha/2H_*^2$. Hence, the frequency can be quadratically tuned by the applied magnetic field.

### 3. Local magnetic field generation

To generate a local magnetic field, a DC current is applied to the two ground lines, as illustrated in the schematic diagram in Fig. 2(a). Here, we investigate the effects of parallel and antiparallel currents flowing through the ground wires, respectively. When antiparallel currents flow through the ground wires, a magnetic field is locally generated on the center conductor of the resonator (Fig. 2(b)). In contrast, the magnetic fields generated by parallel currents in the two ground wires neutralize each other at the center conductor, as depicted in Fig. 2(c). The magnetic field can be calculated using Biot-Savart Law. For a NbN film with a thickness of 22 nm, only the perpendicular component of the magnetic field ($H_z$, in the Z=0 plane) contributes to the frequency tuning [19] and is given by

$$H_z = \frac{I_{dc}}{4\pi b}\left[ln\frac{(y-a)^2}{(y-a-b)^2} \pm ln\frac{(y+a)^2}{(y+a+b)^2}\right], \qquad (7)$$

where $I_{dc}$ is the current flowing through the ground wires. The "+" sign represents antiparallel currents, and the '-' sign denotes parallel currents.

In our experiment, applying a maximum antiparallel current of $I_{dc}$=100 mA results in a 2.5 mT magnetic field at the center conductor (Fig. 2(d)). At a distance of 50 μm from the center conductor, the magnetic field decreases to 0.35 mT (Fig. 2(d)). The magnetic field weakens rapidly with increasing distance from the center conductor. This approach ensures that the locally generated magnetic field has a neglectable impact on other components integrated into the circuits.

## 4. Experimental results and analysis

To characterize the frequency tunability of our resonator, we recorded the forward transmission ($S_{21}$) using a vector network analyzer (VNA) at various ground line currents. The superconducting resonator was mount in a dilution refrigerator operating at 20 mK. The input signal from the VNA was attenuated (-50 dB) to minimize the noise. The output signal, isolated by an isolator, was connected to the VNA through a high electron-mobility-transistor amplifier with a gain of +42 dB. The on-chip input power is -70 dBm. First, we measured $S_{21}$ at 20 mK while appling antiparallel currents (0-100 mA) to the ground wires. As shown in Fig. 3(a), the resonance frequency of the resonator was suppressed with increasing currents, revealing a maximum frequency shift of 54.85 MHz from a 6.21 GHz resonance tone under $I_{dc}$=100 mA.

Although the ground wires are superconducting, the contact resistance between bonding wires and electrodes could induce Joule heating when a significant current is applied. To examine the possible thermal effects on the frequency tuning, we compared the above results to measurements under a parallel-current configuration (Fig. 3(b)) as well as to those obtained by directly varying the temperature (Fig. 3(c)). As shown in Fig. 3(b), the resonance frequency remains relatively stable at low parallel currents, with a sudden shift observed beyond a certain current value (≥ 80 mA). This is in contrast to the effects from antiparallel currents configuration, in which the resonance frequency gradually shifts with currents. Moreover, the frequency shift in the antiparallel-current configuration is much larger than that in the parallel-current configuration. Despite the Joule heating effect being equivalent between the antiparallel and parallel currents, the significantly larger frequency shift in the antiparallel-current configuration suggests that the observed frequency tuning in Fig. 3(a) is dominated by the local magnetic field produced from antiparallel currents in the ground wires.

On the other hand, as depicted in Fig. 3(c), the direct temperature-dependent results reveal a rapid weakening of the resonance signal strength, with a substantial shift (328 MHz) in resonance frequency as the temperature increases from 20 mK to 5 K. When comparing the effect of the antiparallel current configuration (Fig. 3(a)) to the temperature effect (Fig. 3(c)), it is evident that the resonance signal strength from the temperature effect is about 10 dB lower than that from the antiparallel current configuration for a similar frequency shift, as displayed by the dashed lines in Figs. 3(a) and 3(c). This further highlights the advancement of tuning resonance frequency using local magnetic field.

To quantitatively analyze our data, we employ the Diameter Correction Method (DCM) [32] to fit the $S_{21}$ data and extract the resonant frequencies and quality factors. According to the DCM model, the forward transmission can be expressed as:

$$S_{21}(f) = 1 - \frac{Q_L e^{i\phi}}{Q_e} \frac{1}{1 + 2iQ_L \frac{f-f_0}{f_0}}, \qquad (8)$$

where $\phi$ is the asymmetry factor of the resonator, $f_0$ is the resonance frequency, and $Q_L$ is the loaded quality factor, defined as the parallel combination of internal ($Q_i$) and external ($Q_e$) quality factors:

$$\frac{1}{Q_L} = \frac{1}{Q_e} + \frac{1}{Q_i}, \qquad (9)$$

Figures 4(a) and 4(b) respectively display the experimental results (depicted by red lines) of the magnitude and the phase of complex transmission data $S_{21}$ obtained at 20 mK and with zero current in the ground lines. The corresponding fitting lines are shown in blue lines in Figs. 4(a) and 4(b). From the fittings, we obtain the internal quality factor $Q_i \approx 8.2 \times 10^4$ and the zero-field resonant frequency $f_0 \approx 6.21$ GHz. The $Q_i$ value is consistent with those reported previously for resonators fabricated from the same material [Appl. Phys. Lett. 113, 142601 (2018)], revealing the high quality of our NbN film.

Figure 4(c) presents the extracted $I_{dc}$ dependent frequency shift $\Delta f$ for parallel-current (blue) and antiparallel-current (red) configurations. According to Equations (6) and (7), $\Delta f$ induced by the magnetic field $H$ is proportional to $H^2$ and $I_{dc}^2$. As depicted in the inset of Fig. 4(c), $\Delta f$ for the antiparallel-current configuration changes linearly with $I_{dc}^2$, in perfect agreement with the magnetic-field-induced resonance frequency shift mentioned earlier.

As indicated by the blue curve in Fig. 4(c), the $\Delta f$ for the parallel-current configuration remains nearly zero until $I_{dc} \geq 80$ mA. Beyond this point, the corresponding $Q_i$ rapidly decreases and is lower than that for the antiparallel-current configuration (Fig. 4(d)). Notably, the $Q_i$ (red curve in Fig. 4(d)) for the antiparallel-current configuration decreases very slowly for $I_{dc} \geq 60$ mA, and in some instances, it may even slightly increase. This indicates a distinct role of the parallel currents compared to that of the antiparallel currents. In Figs. 4(e) and 4(f), we compare the evolutions of $\Delta f$ and $Q_i$ with parallel currents and those with the temperature. It can be seen that both $\Delta f$ and $Q_i$ curves can be perfectly normalized by $T=\kappa I_{dc}^2$, where $T$ is temperature, and $\kappa$ is a constant. This clearly indicates the frequency tuning in the parallel-current configuration is dominated by the Joule heating-induced temperate change. Figures 4(e) and 4(f) show that the effect of the applied maximal parallel current of $I_{dc}=100$ mA corresponds to an effective temperature of 2.5 K.

Although one might anticipate equivalent Joule heating effects from parallel currents and antiparallel currents, the much larger and slightly increased $Q_i$ with $I_{dc}$ at $I_{dc} \geq 80$ mA (the red curve in Fig. 4(d)) indicates a negligible temperature dependent effect for the antiparallel-current configuration compared to that for the parallel-current configuration. This suggests that parallel currents produce more dissipation than antiparallel currents. Two possible explanations for this effect are: 1) The parallel-current configuration requires current flowing through additional electrodes, thus resulting in larger Joule heating from contact resistance compared to that in the antiparallel-current configuration, as shown in Fig. S1. 2) The parallel currents induce stronger effects on vortex motion in the ground lines, thus generating more dissipation than the antiparallel currents. The real mechanism behind the difference between parallel- and antiparallel-current configurations requires more detailed investigation in the future.

## 5. Conclusion and perspectives

In conclusion, our on-chip method of generating a local magnetic field by applying antiparallel currents to ground wires successfully achieves notable frequency tunability in superconducting resonators. The demonstrated superconducting resonator, featuring an internal quality factor of $8.2 \times 10^4$, a generated magnetic field of 2.5 mT, and a maximum frequency tunability of 54.85 MHz, holds substantial promise for the large-scale integration in quantum circuits. The confined local magnetic field, limited to the resonator's central line, ensures minimal impact on other integrated circuit components.

Our tunable superconducting resonators can benefit a range of applications, such as quantum computing[2-6, 9-11] and single-photon detection[1]. The presence of ground lines that interrupt the ground plane lowers the internal quality factors in comparison to a standard CPW resonator (see Fig. S2). Improvements in grounding, either by reducing resistance or by increasing the capacitance between the ground lines and the external ground plane, could mitigate this issue. Further improving the frequency tuning performance could be possible by fine tuning the resonator's parameters, including materials, geometrical dimensions, and contact resistance. The unique behavior observed in experiments, wherein the frequency tuning effect is predominantly governed by the local magnetic field rather than thermal effects, emphasizes the effectiveness of our methodology. This investigation paves the way for further advancements in superconducting resonators for quantum technologies, highlighting the crucial role of on-chip local magnetic field control in the development of future integrated quantum systems.

**Acknowledgment**

This work is supported by the National Key R&D Program of China (2021YFA0718802 and 2018YFA0209002), the National Natural Science Foundation of China (62274086, 62288101, 61971464, 62101243 and 11961141002), Jiangsu Excellent Young Scholar program (BK20200008 and BK20200060), Jiangsu Outstanding Postdoctoral Program, the Fundamental Research Funds for the Central Universities, and Jiangsu Key Laboratory of Advanced Techniques for Manipulating Electromagnetic Waves.

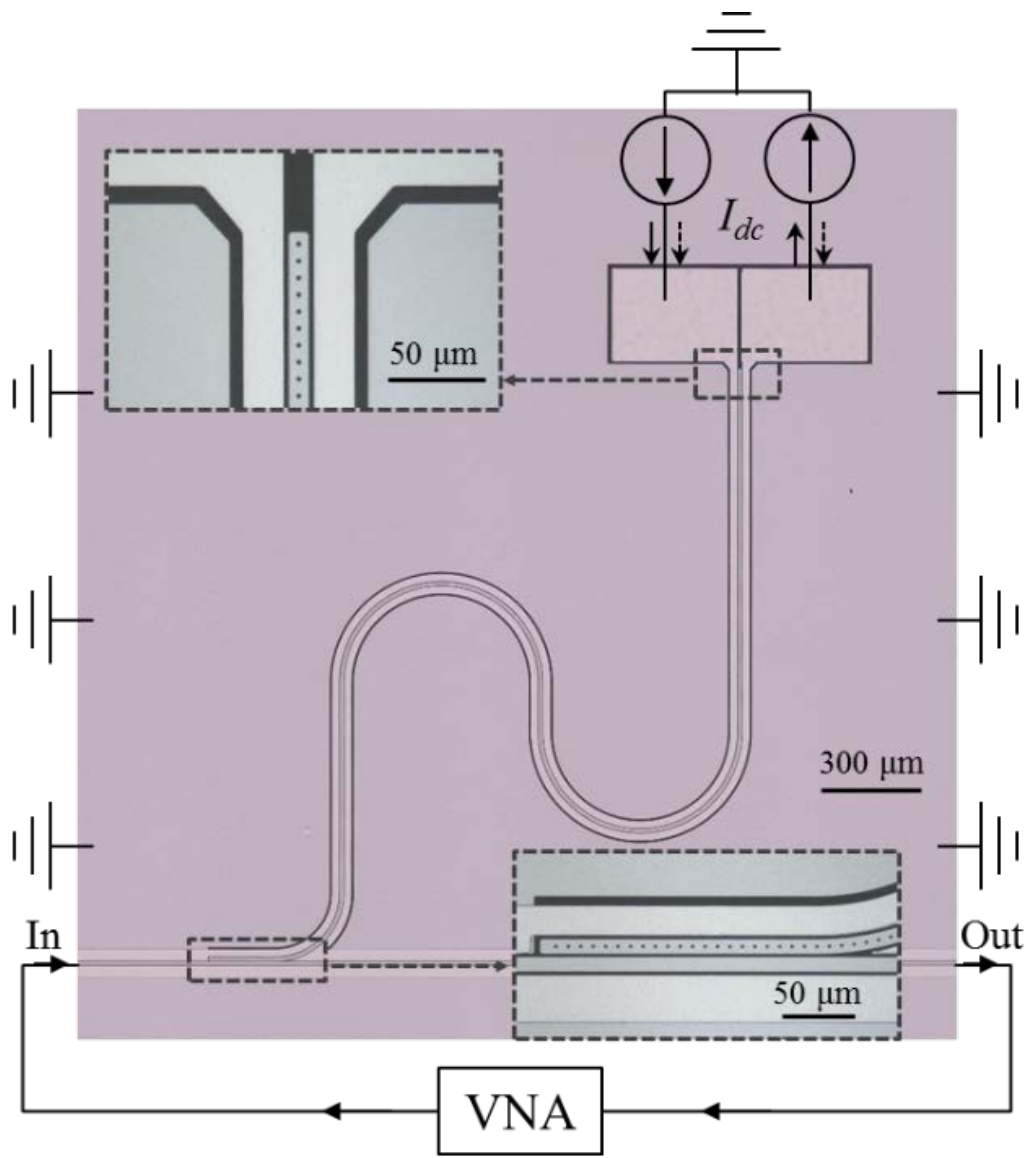

**Fig.1.** Optical photography of the CPW resonator coupled to a feedline. The width and length of the resonator is 10 μm and 4289 μm, respectively. The diameter of the holes in the center conductor is approximately 3 μm. The length of the coupler between the resonator and the feedline is 200 μm.

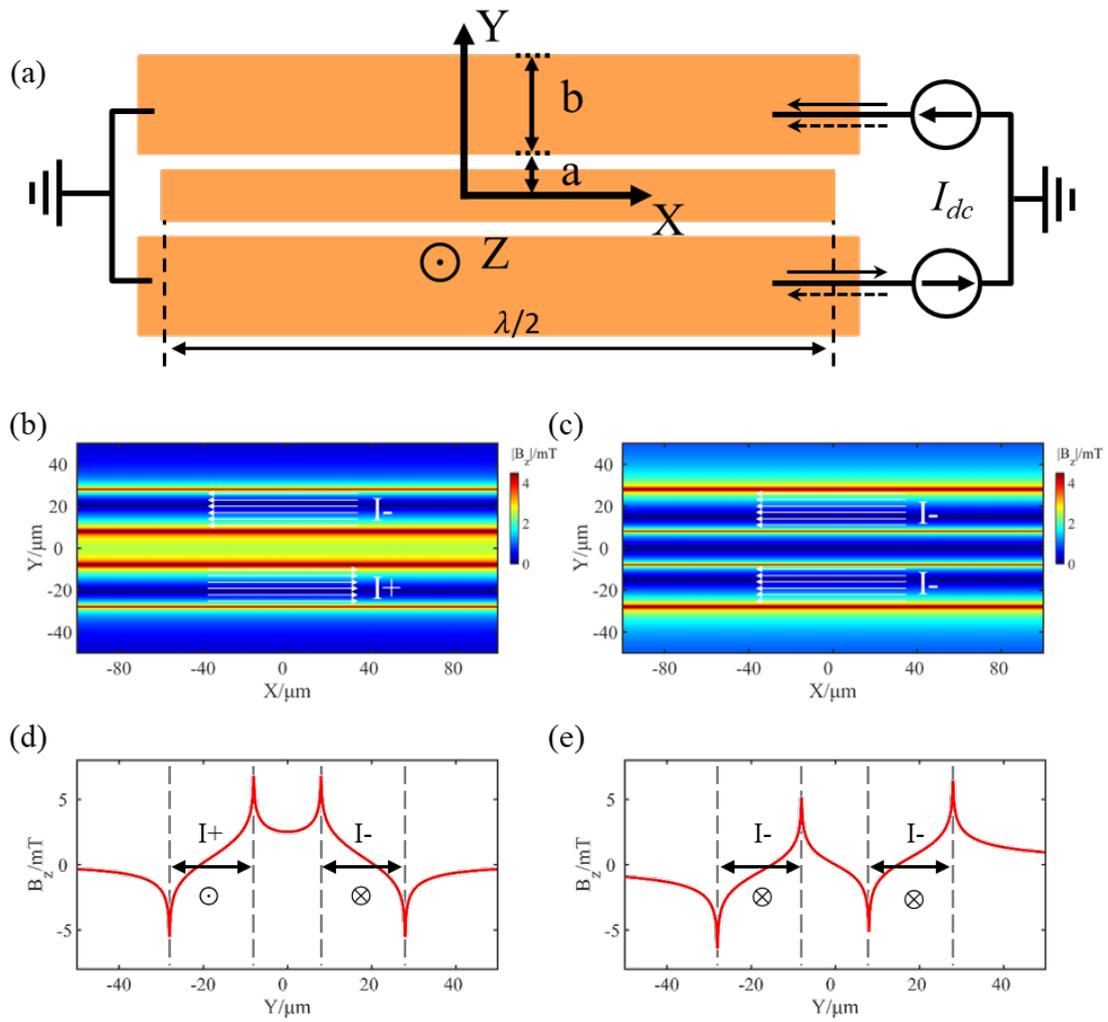

**Fig. 2.** (a) Schematic diagram of the resonator with current applied to the two ground lines. (b and c) Calculated magnetic field distribution on the resonator when applying antiparallel currents (b) and parallel currents (c). (d and e) Magnetic field distribution across the CPW resonator under antiparallel currents (d) and parallel currents (e).

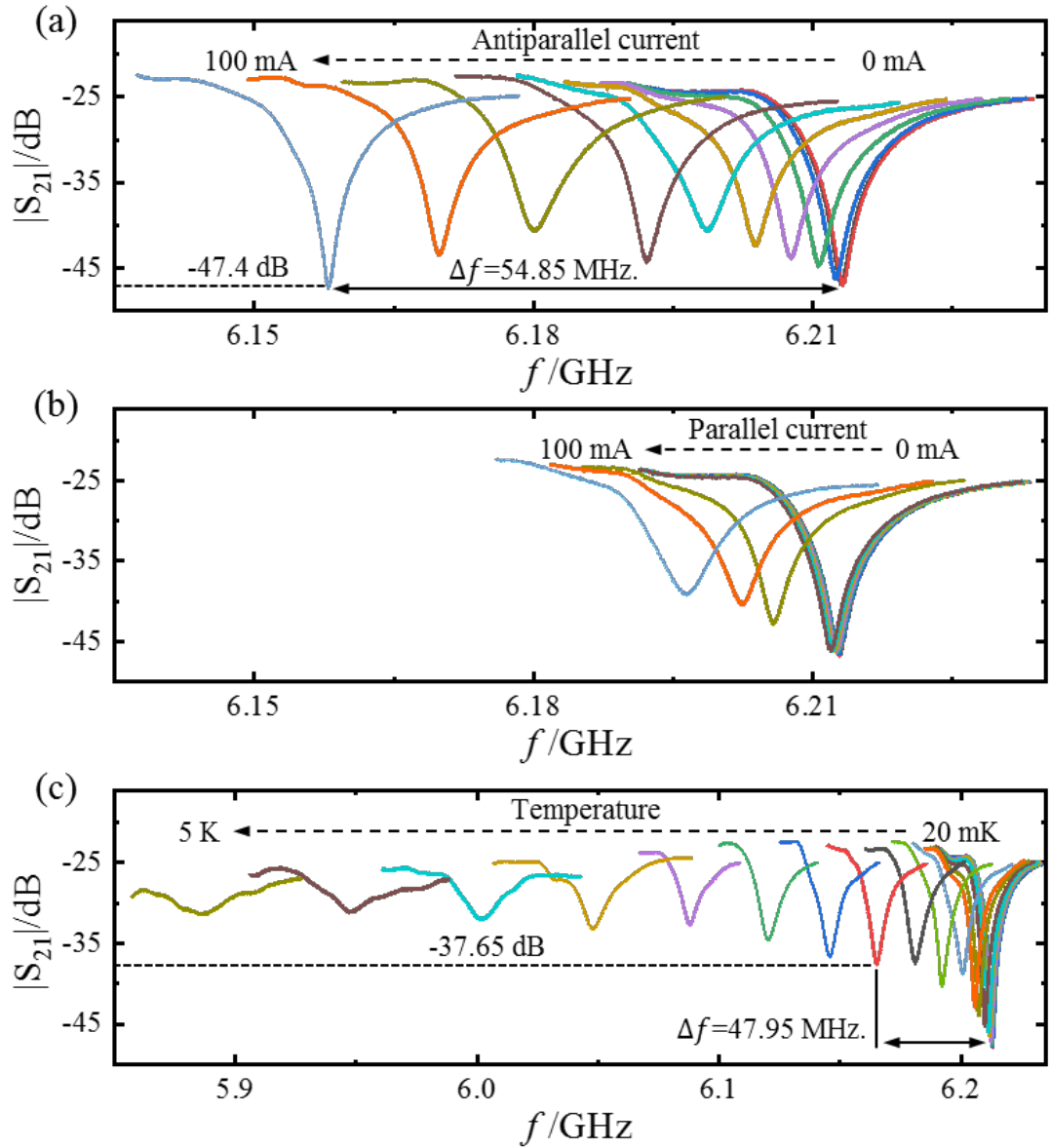

**Fig. 3.** $|S_{21}|$ spectra of the tunable resonator. (a and b) Evolutions of $|S_{21}|$ spectra with antiparallel currents (a) and parallel currents (b), swept from 0 to 100 mA at 20 mK. (c) $|S_{21}|$ spectra as a function of temperature from 20 mK to 5 K.

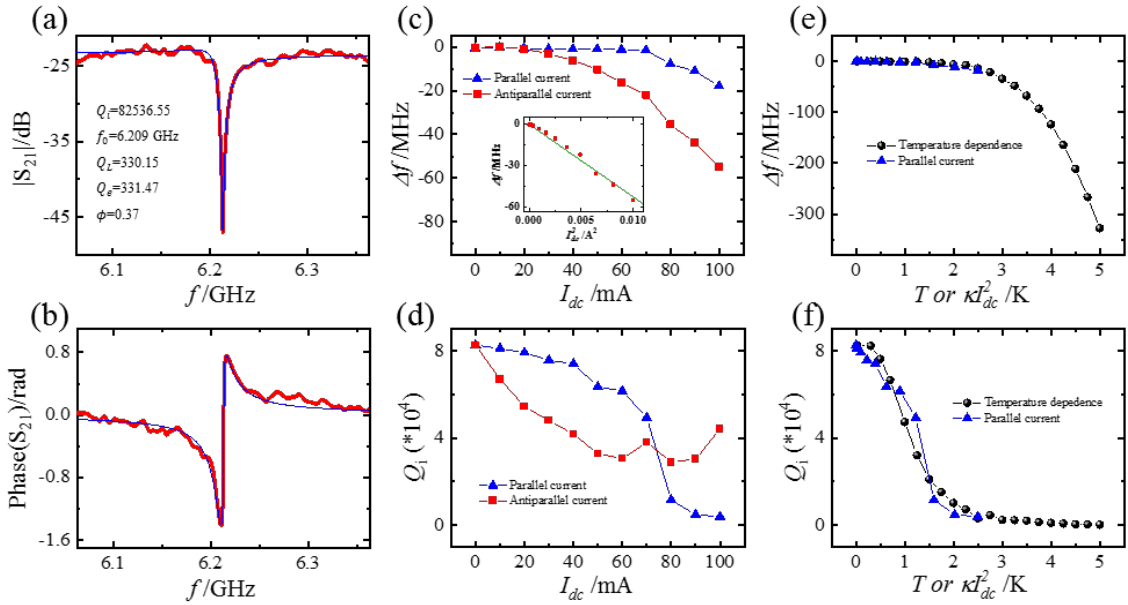

**Fig. 4.** Quantitative analysis of frequency tuning effects. (a and b) The resonance line shapes of experimental data (red) and fitting data (blue) for the magnitude (a) and phase (b) of $S_{21}$. (c and d) The frequency shift $\Delta f$ (c) and the internal quality factor $Q_i$ (d) as a function of parallel currents (blue) and antiparallel currents (red). The inset of (c) is the fitting of $\Delta f$ to Equation (6) for antiparallel-current configuration. (e and f) Comparison between parallel-current dependent (blue) and temperature dependent (black) frequency shift $\Delta f$ (e) and the internal quality factor $Q_i$ (f).

**Supplementary data figure**

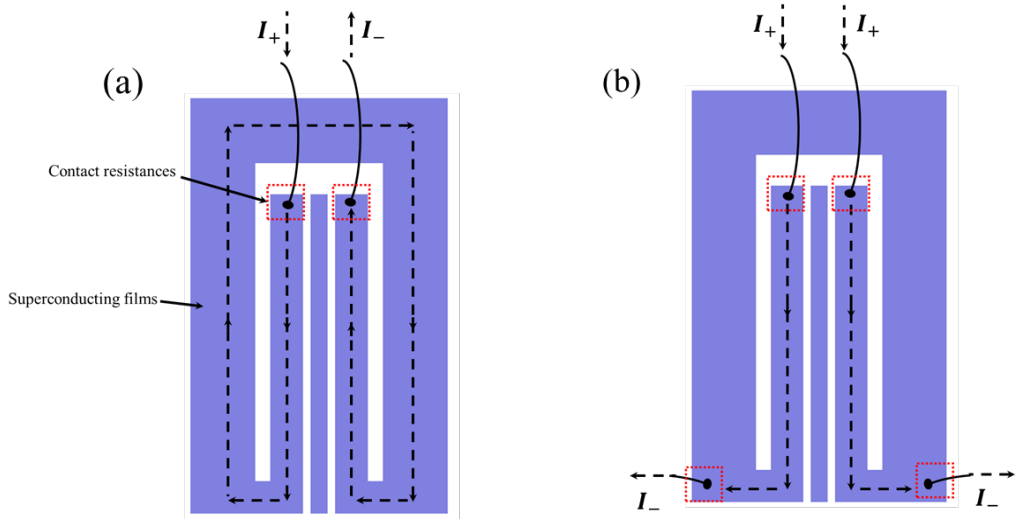

**Fig. S1.** (a) The antiparallel currents path: when applying antiparallel currents, the currents flow from one ground wire to the other through the superconducting ground plane. (b) The parallel currents path: when applying parallel currents, the currents

must flow from the ground plane of the device to that of the PCB through additional wire bonds

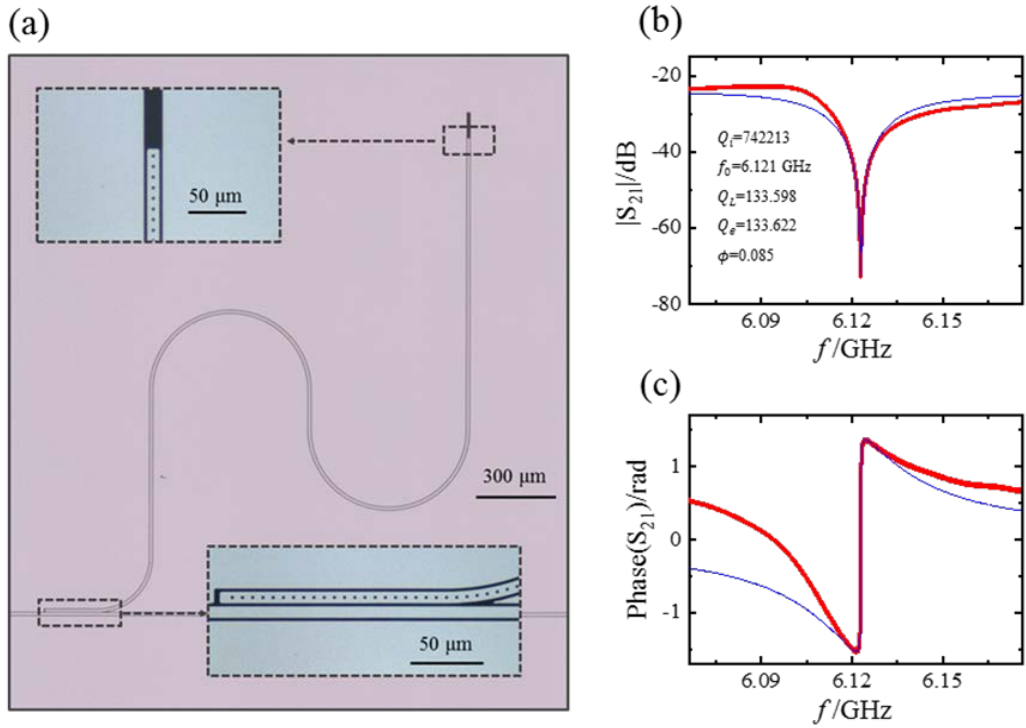

**Fig. S2.** The standard resonator fabricated from the same film as the tunable resonator. (a) Sample image. (b and c) The resonance spectra of experimental data (red) and fitting data (blue) for the magnitude (b) and phase (c) of $S_{21}$.